\documentclass{ws-procs9x6}

\begin{document}

\title{Standard-like models from D-branes}

\author{D. BAILIN}

\address{Centre for Theoretical Physics \\
University of Sussex \\ 
Brighton, BN1 9QJ, UK\\ 
E-mail: D.Bailin@sussex.ac.uk}

\maketitle

\abstracts{I describe the main features of new intersecting D4- and D5-brane  orbifold models that yield the non-supersymmetric standard model, 
up to vector-like matter 
and, in some cases, extra $U(1)$ factors in the gauge group. There are six-stack D4-brane models that have charged-singlet scalar tachyons 
and which either contain all of the Yukawa couplings to the tachyonic Higgs doublets 
that are needed to generate mass terms for the fermions at renormalisable level or possess an unwanted extra $U(1)$ gauge symmetry 
after spontaneous symmetry breaking. In the D5-brane models a minimum of eight stacks is needed. 
%Some of these models are free of unwanted gauged $U(1)$ symmetries and predict ratios of gauge coupling constants equal to the measured values. 
}

\section{Introduction}
The D-brane world offers an attractive, bottom-up route to getting standard-like models from Type II string theory\cite{UTCA}. 
Open strings that begin and end on a stack $a$ of $N_a$ D-branes generate the gauge bosons of the group $U(N_a)$ living in the world volume of the D-branes.
 So the 
standard approach is to start with one stack $a=1$ of three D-branes, another $a=2$ of two D-branes, and $n$ other stacks each having just one D-brane, 
thereby generating 
the gauge group $U(3) \times U(2) \times U(1)^n$. Fermions in bi-fundamental
 representations of the corresponding gauge groups can arise at the intersections of such stacks\cite{BDL}, but to get $D=4$ {\it chiral} fermions 
  the intersecting branes should sit at a singular point in the space transverse to the branes, an orbifold fixed point, for example. In general,
   such configurations yield a non-supersymmetric spectrum, so to avoid the hierarchy problem the
    string scale associated with such models must be no more than a few TeV. Gravitational interactions occur in the bulk ten-dimensional space, and 
    to ensure that the Planck energy has its observed large value, it is necessary that there are large dimensions transverse to the branes\cite{ADD}. 
The D-branes with which we are concerned wrap the 3-space we inhabit and closed 1-, 2- or 3-cycles of a toroidally 
compactified $T^2, \ T^2 \times T^2$ or $T^2 \times T^2 \times T^2$ space. 
Thus getting the correct Planck scale effectively means that only D4- and D5-brane models are viable, 
since for D6-branes there is no dimension transverse to all of the intersecting branes.  

During the past year orientifold models with intersecting D6- and D5-branes
have been constructed that yield 
precisely the fermionic spectrum of the standard model (plus three generations of right-chiral neutrinos)\cite{IMR,CIM}. 
(Other recent work on intersecting brane models, both supersymmetric 
and non-supersymmetric, and their phenomenological implications may be found in\cite{IBM}.)
The spectrum includes 
 $SU(2)_L$ doublet scalar tachyons that may be regarded as the Higgs doublets that break the electroweak symmetry group, but also, unavoidably,
 colour-triplet and charged singlet tachyons either of which is potentially fatal for the phenomenology. In a previous paper\cite{BKL1} 
we studied D4-brane models, transverse to a $Z_3$ orbifold,  having three generations of chiral matter that were constrained to
 have no colour triplet or charged singlet 
scalar tachyons, and which contained the Yukawa couplings to the  Higgs doublets needed to give masses to all quarks and leptons; these models, unavoidably,
also possessed extra, vector-like leptonic matter. The wrapping numbers of the various stacks are constrained by the requirement 
 of Ramond-Ramond tadpole cancellation and also by the requirement that the mixed $U(1)_Y \times SU(2)_L^2$ and $U(1)_Y \times SU(3)_c^2$ anomalies cancel.
  In string theory, these latter constraints derive from a generalised Green-Schwarz mechanism that ensures that the gauge bosons 
  associated with all anomalous $U(1)$s acquire string-scale masses\cite{IRU}. In fact the cancellation of anomalies is necessary, 
  but {\it not} sufficient\cite{IMR}, to
  ensure the masslessness of the gauge boson associated with $U(1)_Y$. 
  When the stronger constraints derived from this Green-Schwarz mechanism are applied we find 
  %We have recently realised 
  that in the models constructed in\cite {BKL1} the 
  weak hypercharge  $U(1)_Y$ survives only as a global, {\it not} a local, symmetry\footnote{A similar observation applies also to 
  the models constructed in\cite{AFIRU2} and Kataoka and Shimojo\cite{IBM}.}. 
  A recent paper\cite{BKL2} was the first to construct semi-realistic intersecting D4-brane orbifold models without this defect. In this talk I 
  outline the features of our new D4-brane models and some results from our attempts to construct analogous models using D5-branes.
   
\section{D4-brane models}
The stacks of D4-branes wrap closed 1-cycles of $T^2$ and are all situated at a fixed point of the transverse $T^2 \times T^2/Z_3$ orbifold.
A stack $a$ is specified by two wrapping numbers $(n_a,m_a)$ that 
specify the number of times $a$ wraps the basis 1-cycles of $T^2$. The generator $\theta$ of the $Z_3$ point group is embedded in the stack of $N_a$ branes as
 $\gamma _{\theta,a} = \alpha ^{p_a}I_{N_a}$, where $\alpha = e^{2\pi i/3}, \ p_a=0,1,2$. 
 The first two stacks $a=1,2$ defined above, that generate a $U(3) \times U(2)$ gauge group, are common to all models; their Chan-Paton 
 factors  $\alpha ^{p_a}$ have $p_1 \neq p_2$.  Besides these we have, in general,  
 three sets $I,J,K$ of $U(1)$ stacks characterised by their Chan-Paton factors: $p_i=p_2  \ \forall i \in I, \ p_1 \neq p_j \neq p_2  \ \forall j \in J$, and 
 $p_k =p_1  \ \forall k \in K$; the sets $I,J,K$ are each divided into two subsets $I_1 \cup I_2=I$ etc., defined so that the weak hypercharge $Y$ is the linear 
 combination of the $U(1)$ charges $Q_a$ 
 \begin{equation}
 -Y=\frac{1}{3}Q_1+\frac{1}{2}Q_2+ \sum_{i_1 \in I_1}Q_{i_1}+ \sum_{j_1 \in J_1}Q_{j_1}+ \sum_{k_1 \in K_1}Q_{k_1}
 \end{equation}
 In general, tachyonic scalars
  arise at intersections between stacks $a$ and $b$ which have the same Chan-Paton factor $p_a=p_b$; the number $I_{ab}=n_am_b-m_an_b$ of 
  intersections gives the number of such particles.
 Thus, Higgs doublets, which are needed to give mass to the fermionic matter, arise at $(2i_1)$ and $(2i_2)$ intersections\footnote{
 In our models, as in all others, the Higgs content is non-minimal; this seems to be a generic feature of models deriving from string 
 theory\cite{GN}.}.
  In\cite{BKL2} we obtained general solutions for the wrapping numbers of three-generation models
  that satisfy the constraints deriving from twisted tadpole 
 cancellation and the requirement that the $U(1)_Y$ gauge boson associated with weak hypercharge does {\it not} get a Green-Schwarz mass. 
 We also demand that there are tree-level Yukawa couplings needed to give masses to all matter. As before\cite{BKL1},
 the three fermion generations include right-chiral neutrinos in all models. At least six 
 stacks are required, and there is a unique, one-parameter family of six-stack models having stacks $1,2,i_1,i_2,j_1,j_2$
  that include the (tachyonic) Higgs doublets needed to generate mass terms for all matter. 
 They have extra vector-like leptonic, but not quark, matter. We find
\begin{equation}
12(L+\bar{L})+6(e^c_L +\bar{e}^c_L)+3(\nu^c_L +\bar{\nu}^c_L)
\end{equation}
Besides 9 Higgs doublets, there are also 3 charged-singlet tachyons, but no colour-triplet tachyons. The gauge coupling strengths 
are inversely proportional to the length of the 1-cycle wrapped by the corresponding stack, so the wrapping numbers determine 
gauge coupling constant ratios, at the string scale. The ratios predicted by the above six-stack models are inconsistent with those measured at the 
electroweak scale, but our previous experience\cite{BKL1} suggests that they might be consistent with only a modest amount of renormalisation 
group running, so they are not necessarily inconsistent with a string scale of a few TeV that is required for  non-superymmetric models such 
as these. It is conceivable too that radiative corrections could render the tree-level, charged-singlet tachyons non-tachyonic.
However, there is a further unwelcome feature of these models that seems unavoidable. It turns out that in the models under discussion 
there is one surviving additional $U(1)$ gauge symmetry, coupled to the observable matter, that is not spontaneously broken at the electroweak 
transition. We can do better if we abandon the requirement of mass terms for all matter at renormalisable level. Then there is a one-parameter family of
 six-stack models 
having stacks $1,2,i_1,j_1,j_2,k_2$ with no 
unwanted (coupled) surviving $U(1)$ gauge symmetry, but which has additional vector-like leptonic and quark matter. We find 
 \begin{equation}
3(d^c_L +\bar{d}^c_L)+6(L+\bar{L})+6(e^c_L +\bar{e}^c_L)+3(\nu^c_L +\bar{\nu}^c_L)
\end{equation}
In addition there are 3 Higgs doublets, 3 charged-singet tachyonic scalars and 3 colour-triplet scalar tachyons. 
 
We have also found two-parameter, eight-stack models having precisely one stack in each of the sets $I_1,I_2,J_1,J_2,K_1,K_2$, 
  that have mass terms at renormalisable level for all matter {\it and} 
 that are free of charged-singlet tachyons. As before, there is 
 extra vector-like leptonic, but not quark, matter. The precise vector-like matter content depends upon one of the parameters. There are several 
 values of this parameter that give gauge coupling constant ratios fairly close to those measured, and one value that comes very close. In this case, 
 therefore, the string scale cannot be far from the electroweak scale. 
 As before\cite{BKL1}, baryon number is anomalous, and survives as a global symmetry, so the proton is stable despite the low string scale. Unfortunately,
  these models too are afflicted with several unwanted $U(1)$ gauge symmetries that survive electroweak spontaneous symmetry breaking.
  
  \section{D5-brane models}
  It is natural to wonder whether similar models, constructed using D5-branes that wrap $T^2 \times T^2$ and are situated at a fixed point of 
  $T^2/Z_3$, can improve upon the results of D4-brane models.  In this case each stack $a$ is characterised by a total of four wrapping numbers 
  $(n_a,m_a)(\tilde{n}_a, \tilde{m}_a)$, two for each wrapped torus. The intersection number is now given by 
  $I_{ab}=(n_am_b-m_an_b)(\tilde{n}_a\tilde{m}_b-\tilde{m}_a\tilde{n}_b)$. 
  Important differences from the D4-brane case arise in connection with the tachyonic scalar states. As before, they arise at intersections of stacks 
  having the same Chan-Paton factor; in fact, only scalar tachyons arise at such intersections in this case. 
  The  difference of the angles $\theta _{ab}$ and $\tilde{\theta} _{ab}$ between the intersecting 1-cycles on the two wrapped tori 
  determine the squared mass of the tachyon at such an intersection, and when they are equal the tachyon is massless. Thus in D5-brane models 
  potentially unwelcome (e.g. charged-singlet and/or colour-triplet) tachyons might be removable by this mechanism. A further difference is that
  tachyons may also arise when 
  the intersecting D5-branes have parallel 1-cycles in one, but not both, of the two wrapped tori\cite{CIM}, 
  i.e. $\theta _{ab}=0$ {\it or} $\tilde{\theta} _{ab}=0$.  The intersection number of the non-parallel 1-cycles determines 
  the number of such states that arise. The distance between the parallel 1-cycles 
  enters the mass formula, so the masses of these states are continuously adjustable.
   When it is zero (or small enough) the states are tachyonic, but when the cycles are far enough apart the states are massive and can be removed from 
   the low-energy spectrum. 
  There are eight constraints on the wrapping numbers deriving from twisted tadpole 
 cancellation and requiring that the $U(1)_Y$ gauge boson associated with weak hypercharge does  not get a Green-Schwarz mass. The 
 constraints are  quadratic in the wrapping numbers, and the only six-stack models that can satisfy them have stacks $1,2,i_1,i_2,k_1,k_2$. 
  Unfortunately, the solution gives $I_{2i_1}=0=I_{2i_2}$ and, since the 1-cycles in both tori are parallel, these models have no Higgs doublets
   (or charged-singlet or colour-triplet tachyons). They are therefore not standard-like and
   we are forced to consider eight-stack models from the outset. 
   
We may obtain non-zero intersection numbers for the Higgs doublets  by taking the sets $I_1$ and $I_2$ to contain more than one stack. 
In particular, if we take each to have two stacks $i_1^{(1,2)}$ and $i_2^{(1,2)}$ respectively, then there are consistent eight-stack, 
three-generation models 
with just $j_1$ (or $k_1$) and $j_2$ (or $k_2$) stacks in addition. A priori, the Green-Schwarz mechanism gives string-scale masses to eight 
linear combinations of the $U(1)$ gauge fields. However, in the models we are considering three of these are lost, and 
 the three surviving massless $U(1)$ gauge symmetries include (an unwanted) one which is coupled to observable matter, but which 
 remains unbroken after electroweak spontaneous symmetry breaking. We have also investigated models in which the Higgs doublets 
 arise from intersections on just one of the wrapped tori, the 1-cycles on the other being parallel.  We find eight-stack, three-generation models 
 satisfying the constraints with just one stack in each of the sets $I_1,I_2,J_1,J_2,K_1,K_2$, 
 and some of these are free of unwanted coupled U(1) gauge symmetries. The models have extra vector-like 
 leptonic, but not quark, matter, and charged-singlet scalar tachyons. The 
 two parameters may be chosen so that the tree-level gauge coupling constant ratios are equal to those measured, 
 consistent with a string scale very close to the electroweak scale.
 These models will be analysed elsewhere\cite{BKL3}.
 
\section*{Acknowledgements}
It is a pleasure to 
acknowledge an enjoyable collaboration with George Kraniotis and Alex Love with whom all of this work was done.
 This account is a modified version of the 
talk I gave in Oxford in July, 2002. The research is supported in part by PPARC and by the German-Israeli 
Foundation for Scientific Research (GIF).

\end{document}